\documentclass[10pt,letterpaper]{article}

\usepackage[T1]{fontenc}
\usepackage[utf8]{inputenc}
\usepackage{mathptmx}  
\usepackage{microtype}
\usepackage[letterpaper,margin=0.8in,columnsep=0.25in]{geometry}

\usepackage{amsmath}
\usepackage{siunitx}
\usepackage{graphicx}
\graphicspath{{figures/}}
\usepackage{booktabs}
\usepackage{tabularx}
\usepackage{array}
\usepackage[font=small,labelfont=bf,labelsep=space]{caption}
\usepackage{stfloats}            
\usepackage{xcolor}
\usepackage{hyperref}  
\hypersetup{
  colorlinks = true,   
  linkcolor  = blue,   
  citecolor  = blue,   
  urlcolor   = blue    
}
\usepackage{xurl}   
\usepackage{authblk}

\usepackage{orcidlink}

\usepackage[round,authoryear,sort]{natbib}
\newcommand{\doi}[1]{\href{https://doi.org/#1}{doi:\nolinkurl{#1}}}

\usepackage[acronym,nomain,nonumberlist,nopostdot,nogroupskip,toc=false]{glossaries}
\setacronymstyle{long-short}
\makenoidxglossaries

\newacronym{apt}{APT}{atom probe tomography}
\newacronym{crism}{CRISM}{Compact Reconnaissance Imaging Spectrometer for Mars}
\newacronym{dragns}{DraGNS}{Dragonfly Gamma-Ray and Neutron Spectrometer}
\newacronym{duv}{DUV}{deep-ultraviolet}
\newacronym{fib}{FIB}{focused ion beam}
\newacronym{fire}{FIRE}{Field Illumination with Radioactive Elements}
\newacronym{ftir}{FTIR}{Fourier transform infrared}
\newacronym{gcms}{GC-MS}{gas chromatography-mass spectrometry}
\newacronym{hpge}{HPGe}{high-purity germanium}
\newacronym{ica}{ICA}{independent component analysis}
\newacronym{libs}{LIBS}{laser-induced breakdown spectroscopy}
\newacronym{microct}{micro-CT}{X-ray micro-computed tomography}
\newacronym{msr}{MSR}{Mars Sample Return}
\newacronym{naa}{NAA}{neutron activation analysis}
\newacronym{nanosims}{NanoSIMS}{nanoscale secondary ion mass spectrometry}
\newacronym{pgnaa}{PGNAA}{prompt gamma neutron activation analysis}
\newacronym{pixl}{PIXL}{Planetary Instrument for X-ray Lithochemistry}
\newacronym{pls}{PLS}{partial least squares}
\newacronym{sam}{SAM}{Sample Analysis at Mars}
\newacronym{sherloc}{SHERLOC}{Scanning Habitable Environments with Raman and Luminescence for Organics and Chemicals}
\newacronym{sims}{SIMS}{secondary ion mass spectrometry}
\newacronym{tims}{TIMS}{thermal ionization mass spectrometry}
\newacronym{xanes}{XANES}{X-ray absorption near-edge structure}
\newacronym{xrf}{XRF}{X-ray fluorescence}

\usepackage{titlesec}
\titleformat{\section}{\normalfont\large\bfseries\raggedright}{\thesection.}{0.5em}{}
\titleformat{\subsection}{\normalfont\normalsize\bfseries\raggedright}{\thesubsection}{0.5em}{}
\titlespacing*{\section}{0pt}{10pt plus 2pt}{4pt}
\titlespacing*{\subsection}{0pt}{8pt plus 2pt}{3pt}

\hypersetup{
  pdftitle    = {Analytical Chemistry at the Planetary Frontier},
  pdfauthor   = {Mavia Anjum},
  pdfkeywords = {planetary analytical chemistry, remote sensing, LIBS, X-ray fluorescence, sample return, chemometrics}
}

\title{\bfseries Analytical Chemistry at the Planetary Frontier}
\author{Mavia Anjum\,\orcidlink{0009-0005-2773-4172}}
\affil{Department of Physics, University of Idaho, Moscow, ID 83844, USA}
\date{}

\begin{document}

\maketitle
\thispagestyle{plain}

\vspace{-1.5em}
\begin{center}
\small
\textbf{Correspondence:} \href{mailto:manjum@uidaho.edu}{manjum@uidaho.edu},
\href{mailto:mav.mavia14@gmail.com}{mav.mavia14@gmail.com}\\[2pt]
\textbf{ORCID:} \href{https://orcid.org/0009-0005-2773-4172}{0009-0005-2773-4172}
\end{center}

\vspace{1em}
\begin{center}\textbf{\large Abstract}\end{center}
\vspace{-0.5em}
\noindent
Planetary science depends on chemical measurements that are made under conditions very different
from those in a terrestrial laboratory. This review treats the analytical methods of planetary
science as one connected chain. The chain starts with orbital remote sensing, continues through
in-situ measurements by landers and rovers and the collection and curation of samples, and ends
with the micro-analysis of returned material on Earth. Orbital methods, including visible and
near-infrared reflectance, thermal emission and gamma-ray and neutron spectroscopy, are discussed
first. In-situ techniques are then reviewed, including \gls{libs}, micro-focus \gls{xrf},
deep-ultraviolet Raman and fluorescence spectroscopy, evolved gas analysis with \gls{gcms} and
miniaturized mass spectrometry. The contamination control and curation practices that protect
returned samples are described, followed by high-precision laboratory techniques such as
\gls{sims} and NanoSIMS, \gls{apt}, \gls{tims}, synchrotron micro-XRF and \gls{xanes}
spectroscopy. Accuracy, calibration and reference materials are discussed as the link between
in-situ and laboratory data, and the role of chemometrics and data fusion in combining these
datasets is examined. Flight and laboratory instruments are compared in terms of spatial
resolution, detection limit, sample consumption and throughput. Active neutron-gamma spectroscopy
planned for the Dragonfly mission to Titan is also discussed. This comparison shows that in-situ
and laboratory techniques are complementary. In-situ instruments provide context and select
samples, while laboratory techniques provide isotopic, molecular and speciation data that cannot
yet be obtained in situ.

\vspace{0.5em}

\glsresetall

\clearpage
\twocolumn

\section{Introduction}

Chemical analysis is central to planetary science. Questions about the habitability of ancient
Mars, the origin of water and organic molecules on the early Earth and the age of the Solar System
are all answered through the elemental, isotopic and molecular composition of specific materials
\citep{seaton2022}. The reliability of each answer therefore depends on the quality of the
measurement behind it. Planetary materials are either located on other bodies or reach Earth in
very small amounts, often in milligram or microgram quantities. Instruments sent to other planets
must work in vacuum or in thin atmospheres, survive large temperature changes and operate within
strict limits of mass, power and data volume. In addition, a measurement on a pristine planetary
surface usually cannot be repeated. These conditions make planetary analytical chemistry different
from routine laboratory analysis, even though the physics of the techniques is the same. The Mars
2020 mission is a good example of the modern approach. Its Perseverance rover carries a set of
analytical instruments to Jezero crater and also collects and seals rock and regolith samples for
possible return to Earth \citep{farley2020}.

\begin{figure*}[t]
  \centering
  \includegraphics[width=0.92\textwidth]{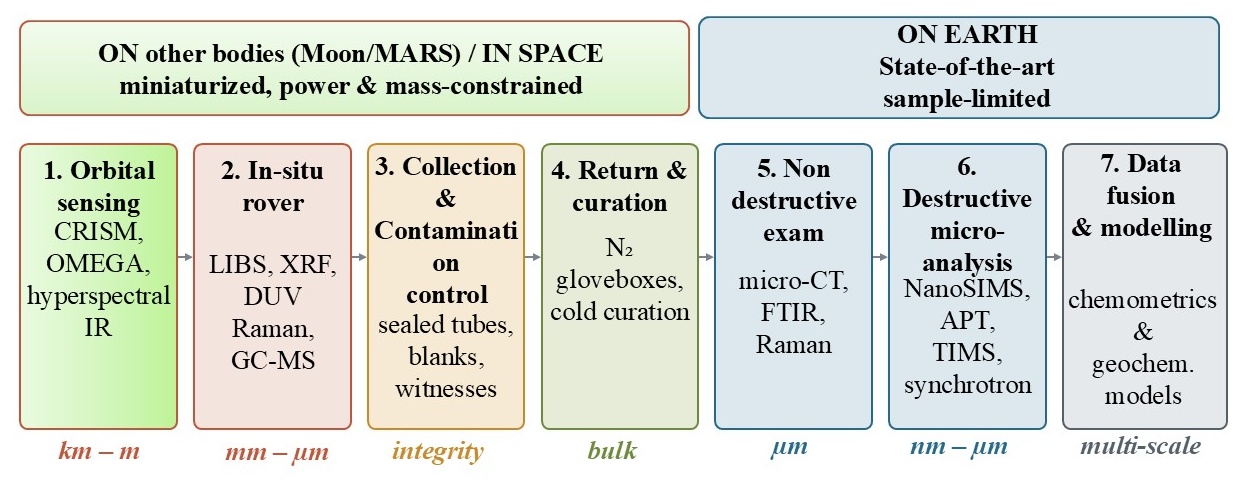}
  \caption{The planetary analytical chain, from orbital remote sensing to laboratory
  micro-analysis. Spatial resolution and analytical detail increase along the chain, and each step
  must preserve the chemical state of the sample established in the previous step. Instruments
  used on Mars or in space are small and limited in power, while laboratory instruments on Earth
  offer the highest performance but are limited by the small mass of returned material}
  \label{fig:chain}
\end{figure*}

Most techniques used in planetary science are also used in terrestrial laboratories. These include
atomic emission spectroscopy, X-ray fluorescence, mass spectrometry and vibrational spectroscopy.
The statistical tools used to process their data, such as principal component analysis and
multivariate calibration, are also common in environmental and radioanalytical studies.
\citet{seaton2022} reviewed the analytical instruments flown on planetary missions over the last
two decades, and \citet{seaton2021} reviewed their use in the search for life. However, these
techniques are usually discussed separately. In practice they form a chain in which each
measurement depends on the steps before it. For example, the oxidation state of iron in a returned
grain is meaningful only if it did not change during collection, transport and storage. Similarly,
a small isotopic anomaly can be interpreted only if terrestrial contamination has been measured and
controlled. There has been no review that follows a sample through the whole chain, from orbital
observation to atom probe analysis, and discusses the analytical limits and calibration problems
at each step. The objective of this review is to address this gap. The analytical chain is shown
in Fig.~\ref{fig:chain}. This review also compares what can be measured in situ with what can be
measured after a sample reaches a laboratory on Earth.

This chain has already produced important results. \citet{eigenbrode2018} reported organic matter
preserved in 3-billion-year-old mudstones at Gale crater using the Curiosity rover. Laboratory
studies showed that samples returned from asteroid Ryugu are similar to the most primitive
carbonaceous chondrites \citep{yokoyama2023}, and that the Bennu samples returned by OSIRIS-REx are
rich in carbon and hydrated minerals \citep{lauretta2024}. The review is organized as follows.
Section~\ref{sec:orbital} discusses orbital and remote methods. Section~\ref{sec:insitu} reviews
in-situ instruments. Section~\ref{sec:contam} describes contamination control and sample curation.
Section~\ref{sec:lab} discusses high-precision laboratory techniques. Section~\ref{sec:accuracy}
considers accuracy, calibration and reference materials. Section~\ref{sec:chemometrics} describes
chemometrics and data fusion. Section~\ref{sec:comparison} compares in-situ and laboratory
analysis. Section~\ref{sec:titan} discusses active neutron-gamma spectroscopy for Titan, and
Section~\ref{sec:conclusions} presents the conclusions. The focus throughout is on analytical
chemistry, including sampling, calibration, sensitivity and data processing, rather than on
geological interpretation.

\section{Orbital and Remote Compositional Analysis}\label{sec:orbital}

The analytical chain starts before any spacecraft lands. Orbital instruments map surface
composition over large areas, provide regional context for local measurements and help select
landing and sampling sites \citep{seaton2022}. Visible and near-infrared reflectance spectroscopy is
widely used for mineral mapping. The \gls{crism} on the Mars Reconnaissance Orbiter measures
reflected light from about \SIrange{0.36}{3.9}{\micro\metre} and reaches a spatial resolution of
about \SI{18}{\metre} per pixel in its targeted mode \citep{murchie2007}. Minerals are identified
from their absorption features, such as the water and hydroxyl bands of clay minerals, the bands of
carbonates and sulfates and the crystal field bands of iron. These features show where liquid water
once altered the Martian crust. Orbital detection of clay minerals and carbonates in the delta of
Jezero crater was one of the main reasons it was selected as the landing site of Perseverance
\citep{farley2020}.

Thermal emission spectroscopy extends mineral mapping into the mid-infrared region. In this region
the vibrational bands of silicate minerals combine in an approximately linear way. Hence, the
mineral composition of a surface can be estimated by fitting a measured spectrum as a combination
of reference mineral spectra. \citet{wyatt2001} used this approach on terrestrial and Martian
volcanic rocks to determine their mineralogy, chemistry and classification. This spectral unmixing
is mathematically similar to the multivariate calibration methods discussed in
Section~\ref{sec:chemometrics}.

Elemental composition from orbit is measured by gamma-ray and neutron spectroscopy. In the passive
method, galactic cosmic rays strike the surface and produce neutrons and gamma rays. The energies
and fluxes of these emissions depend on the elements present. The Gamma-Ray Spectrometer suite on
Mars Odyssey used this method to map hydrogen, which indicates water and ice, as well as elements
such as potassium, thorium, iron and chlorine over the whole planet \citep{boynton2004}. The
spatial resolution of this method is coarse, of the order of hundreds of kilometers. This passive
technique is the basis of the active neutron-gamma method discussed in Section~\ref{sec:titan}. On
a body with a thick atmosphere, cosmic rays cannot reach the surface, and the instrument must
provide its own neutrons.

Remote methods cover large areas but have lower spatial resolution and less chemical detail than
in-situ and laboratory methods. Their main analytical challenge is converting a measured spectrum
or particle flux into composition. This step depends on calibration and modeling, and any error
affects later decisions such as the choice of landing site and samples. Nevertheless, remote
methods provide the context in which more sensitive but local measurements are placed.

\section{In-Situ Planetary Analytical Instrumentation}\label{sec:insitu}

In-situ instruments analyze samples where they are found. The surfaces are usually unprepared and
often covered with dust. The instruments must convert spectra into composition with little human
intervention and under conditions that cannot be fully reproduced during calibration on Earth. For
this reason, most in-situ instruments use optical or mass spectrometric methods that need little or
no sample preparation and that fit within the available mass and power. On Mars, most of the
analytical work is done by four groups of techniques. These are atomic emission and \gls{xrf}
spectroscopy for elemental composition, deep-ultraviolet Raman and fluorescence spectroscopy for
minerals and organic compounds, evolved gas analysis with \gls{gcms} for volatiles and organic
molecules, and miniaturized mass spectrometry.

\subsection{Laser-induced breakdown spectroscopy and X-ray fluorescence}\label{sec:libs}

In \gls{libs}, a pulsed laser is focused on a target to produce a small plasma. The atomic emission
spectrum of the plasma gives the elemental composition of the ablated material. ChemCam on the
Curiosity rover was the first \gls{libs} instrument used on another planet \citep{wiens2012}. It
analyzes targets at distances of up to about \SI{7}{\metre} with a laser spot of about
\SIrange{350}{550}{\micro\metre}. Its successor, SuperCam on the Perseverance rover, combines
\gls{libs} with time-resolved Raman and luminescence spectroscopy and with visible and infrared
reflectance spectroscopy \citep{wiens2021,maurice2021}. This allows elemental and mineralogical
data to be collected from the same target. The main difficulty in planetary \gls{libs} is
quantification. Under the assumption of local thermodynamic equilibrium, the intensity of an
emission line is given by Eq.~(\ref{eq:libs}).
\begin{equation}
  I_{ki} = F\,n\,\frac{g_k A_{ki}}{U(T)}\,\exp\!\left(-\frac{E_k}{k_B T}\right)
  \label{eq:libs}
\end{equation}
Here $F$ is a factor that depends on the instrument and the plasma geometry, $n$ is the number
density of the emitting species, $g_k$ and $A_{ki}$ are the statistical weight and transition
probability of the upper level, $U(T)$ is the partition function, $E_k$ is the energy of the upper
level, $k_B$ is the Boltzmann constant and $T$ is the plasma temperature. The plasma temperature
depends on the composition and physical state of the target. Therefore, the intensity of an
emission line depends not only on the concentration of that element but also on the rest of the
sample. This is known as the chemical matrix effect. Due to this effect, simple univariate
calibration curves are not suitable for the wide range of rock compositions found on a planetary
surface. Instead, the ChemCam team built multivariate calibration models from a large set of
geochemical standards analyzed under Martian atmospheric conditions. \citet{clegg2017} recalibrated
ChemCam with an expanded database of several hundred standards and combined \gls{pls} regression
with \gls{ica}. \citet{boucher2015} compared several machine learning regression methods for the
\gls{libs} analysis of major elements in rocks.

\Gls{xrf} provides an independent measurement of elemental composition. \Gls{xrf} is generally
easier to quantify than \gls{libs} because the intensity of a characteristic X-ray line depends on
known photoionization cross sections and matrix absorption, not on the temperature of a plasma. The
\gls{pixl} on the robotic arm of Perseverance focuses an X-ray beam of about
\SI{120}{\micro\metre} onto a rock surface \citep{allwood2020}. The beam is scanned across the
surface to produce elemental maps of thousands of points, which are matched with close-up images.
In this way, the chemistry of individual grains can be related to the rock texture without sample
preparation. \Gls{libs} and \gls{xrf} are therefore complementary. \Gls{libs} can analyze many
targets quickly from a distance, while \gls{pixl} gives detailed maps of selected targets within
reach of the arm.

\subsection{Deep-ultraviolet Raman and fluorescence spectroscopy}\label{sec:sherloc}

Organic compounds on a planetary surface must be detected with a method that is sensitive to
molecular structure, gives spatial information and does not require sample handling that could add
terrestrial organic contamination. The \gls{sherloc} instrument on Perseverance uses \gls{duv}
Raman and fluorescence spectroscopy for this purpose \citep{bhartia2021}. It uses a
\SI{248.6}{\nano\metre} laser focused to a spot of about \SI{100}{\micro\metre}, which is scanned
across the target to produce chemical maps that are matched with microscopic images. \Gls{duv}
excitation has two advantages. First, fluorescence from aromatic organic compounds appears at
longer wavelengths than the Raman signal, so the weak Raman bands are not hidden by fluorescence as
they often are with visible excitation. Second, when the laser energy is close to an electronic
transition of the molecule, resonance enhancement greatly increases the Raman signal of aromatic
compounds, so that trace organics can be detected. Reliable identification requires careful
wavelength and background calibration under flight conditions. \citet{uckert2021} described this
calibration for \gls{sherloc}. Since \gls{sherloc} does not touch or destroy the sample, it can be
used both to screen targets for organic compounds and to record the organic and mineral context of
samples before they are sealed.

\subsection{Evolved gas analysis and gas chromatography-mass spectrometry}\label{sec:sam}

\Gls{libs}, \gls{xrf} and Raman analyze solids with light. In contrast, the \gls{sam} suite on
Curiosity analyzes the gases released from drilled powder \citep{mahaffy2012}. The powder is heated
in an oven, and the released gases are analyzed by evolved gas analysis, gas chromatography, a
quadrupole mass spectrometer and a tunable laser spectrometer. \Gls{sam} also has a wet chemistry
capability to derivatize polar compounds that cannot be analyzed directly by pyrolysis. \Gls{sam}
can measure atmospheric composition and the isotope ratios of light elements, and it can search
for organic molecules in rocks and soil. \citet{eigenbrode2018} used \gls{sam} to detect
thiophenic, aromatic and aliphatic compounds released from 3-billion-year-old mudstones in Gale
crater and suggested that sulfurization helped preserve the organic matter. This result also shows
the limits of in-situ organic analysis. Pyrolysis breaks large molecules into fragments, can cause
reactions between organic matter and oxychlorine compounds in the soil, and adds background signals
from the instrument. Hence, the organic composition obtained in situ is incomplete, and detailed
molecular structure is difficult to determine.

\subsection{Miniaturized mass spectrometry for landed platforms}\label{sec:ms}

Mass spectrometry is the most general in-situ technique and can measure elemental, isotopic and
molecular composition with little sample preparation. \citet{wurz2012} reviewed its use in
planetary science and noted that mass spectrometers have been used to study the atmospheres of
Mars, Venus, Jupiter, Saturn and Titan and in missions to comets. For solid surfaces, compact laser
ablation and laser desorption time-of-flight mass spectrometers have been developed for landers
\citep{wurz2012}. These instruments measure the elemental and isotopic composition of solid samples
without preparation and at the micrometer scale. Their small size makes them useful for missions
where sample return is not possible. However, laser ionization can cause elemental fractionation,
which is similar to the matrix effect in \gls{libs}. This effect must be corrected by calibration
with matrix-matched standards. In-situ mass spectrometry therefore lies between the survey
instruments described above and the high-precision laboratory mass spectrometry discussed in
Section~\ref{sec:lab}.

\section{Contamination Control and Sample Curation}\label{sec:contam}

\subsection{Blanks, witness samples and cleanliness requirements}

Samples brought to Earth are useful only if their own chemistry can be separated from contamination
added by the spacecraft, the drill, the sample tube or the curation facility. The main problem is
therefore not detection but discrimination. This is the same principle as blank correction in
terrestrial trace analysis, but the required cleanliness is much stricter. Two approaches are
used. Contamination control reduces the amount of terrestrial material introduced into the
samples. Contamination knowledge measures the contamination that cannot be removed, so that it can
be subtracted or taken into account later. The Mars 2020 mission uses both approaches. It set
strict organic and inorganic cleanliness requirements for the sample tubes and related hardware. It
also carries witness tubes that are opened on Mars without collecting rock, so that the
contamination environment of the rover can be measured later in the laboratory \citep{farley2020}.
\citet{maltais2023} described the cleanliness requirements, precision cleaning, processing and
verification of the sample tubes and showed that meeting these requirements was a major
manufacturing and measurement challenge. Biological contamination is also controlled because a
single terrestrial organism in a returned sample could affect the search for life.
\citet{chen2023} described how the biological cleanliness of the Mars 2020 returned sample hardware
was controlled and verified.

\begin{figure*}[t]
  \centering
  \includegraphics[width=0.82\textwidth]{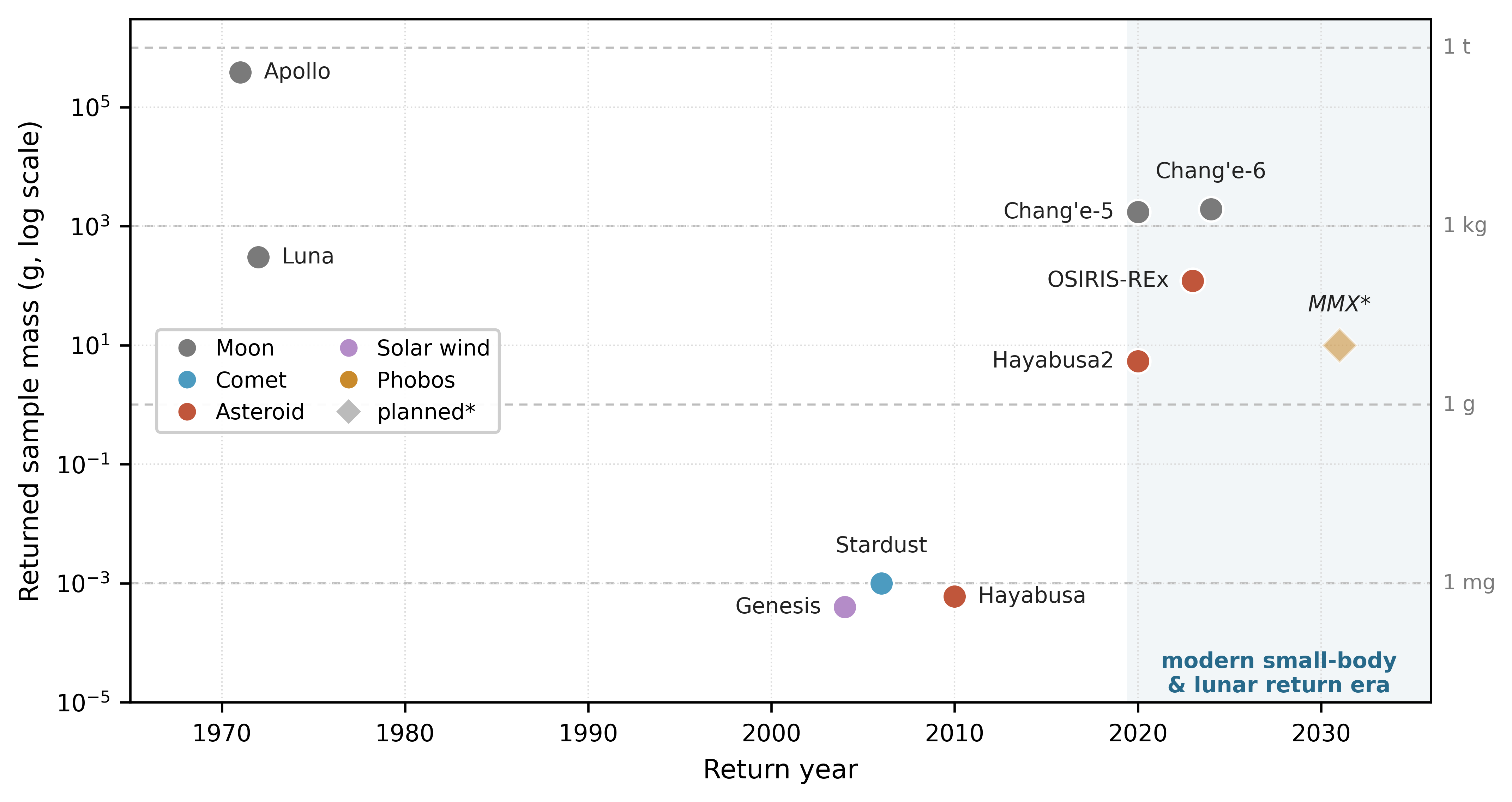}
  \caption{Returned sample mass of extraterrestrial sample return missions plotted against return
  year. Luna represents the combined return of Luna 16, 20 and 24, and the diamond marks the
  planned Martian Moons eXploration (MMX) return}
  \label{fig:missions}
\end{figure*}

The mass of material available for laboratory analysis varies widely between missions, as shown in
Fig.~\ref{fig:missions}. Apollo returned about \SI{382}{\kilo\gram} of lunar material, while
Genesis, Stardust and Hayabusa each returned about \SI{1}{\milli\gram} or less. Hayabusa2 returned
\SI{5.4}{\gram} of material from Ryugu \citep{yokoyama2023}, and OSIRIS-REx returned
\SI{121.6}{\gram} from Bennu \citep{lauretta2024}. For these small-body missions, the scientific
output depends mainly on how much information can be extracted from each grain. The samples
collected by Perseverance are still on Mars. The NASA \gls{msr} program was terminated in the
fiscal year 2026 budget in January 2026, and at present there is no approved mission to bring
these samples to Earth \citep{planetarysociety2026}.

\subsection{Non-destructive preliminary examination}

When a returned sample container is opened, the analyses are carried out from the least to the
most destructive, so that all information that can be obtained without changing the sample is
collected first. Preliminary examination therefore uses non-destructive methods. \Gls{microct}
maps the three-dimensional structure, porosity and dense phases. \Gls{ftir} and Raman
micro-spectroscopy identify hydrated minerals, carbonates and organic functional groups. Optical
and electron microscopy record the texture. Contrast in \gls{microct} images comes from the
attenuation of X-rays passing through the sample, which follows the Beer-Lambert law given in
Eq.~(\ref{eq:beer}).
\begin{equation}
  I = I_{0}\exp\!\left(-\frac{\mu}{\rho}\,\rho\,x\right)
  \label{eq:beer}
\end{equation}
Here $I$ and $I_0$ are the transmitted and incident intensities, $\mu/\rho$ is the mass
attenuation coefficient, $\rho$ is the density and $x$ is the path length. A tomographic
reconstruction is therefore a three-dimensional map of the product of density and mass attenuation
coefficient. The same relation is used in radiation shielding calculations. \citet{nakamura2023}
applied this staged approach to Ryugu samples. They first imaged the particles by synchrotron X-ray
computed tomography and then applied progressively more destructive analyses, so that the maximum
information was obtained from each particle. The Bennu samples returned by OSIRIS-REx were handled
in nitrogen-filled gloveboxes, and their bulk physical, chemical and mineral properties were
measured before the samples were distributed to research teams \citep{lauretta2024}. In both cases,
non-destructive examination was an important analytical step in its own right because it records
the three-dimensional context and volatile-bearing phases that would be lost during later
destructive analysis.

\section{High-Precision Laboratory Micro-Analysis}\label{sec:lab}

\begin{figure*}[t]
  \centering
  \includegraphics[width=0.62\textwidth]{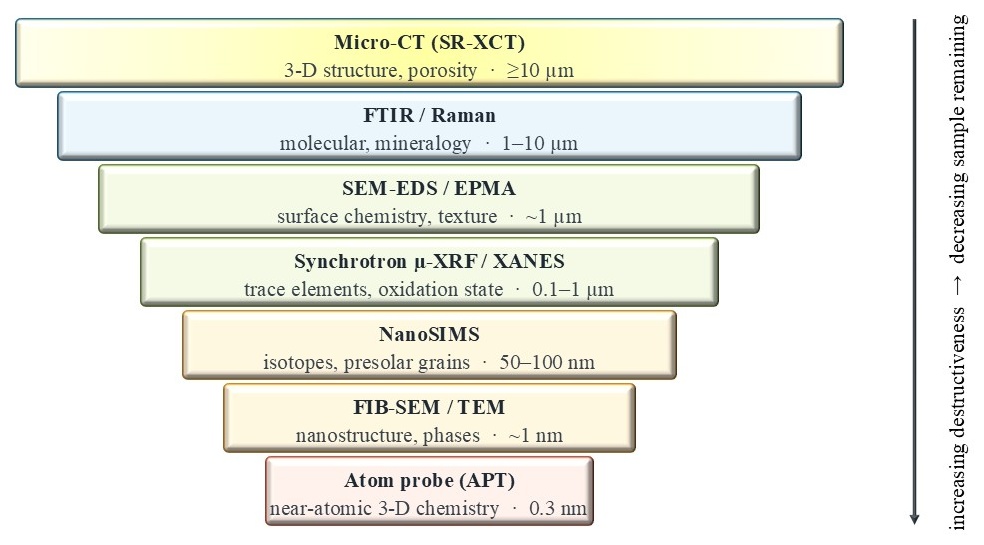}
  \caption{A typical coordinated micro-analysis sequence for a single sub-milligram grain,
  arranged from least to most destructive. Non-destructive tomography and vibrational spectroscopy
  are carried out first, followed by surface chemical and speciation mapping, and finally the
  isotopic and near-atomic analyses that consume the sample. This order increases the total
  information obtained from a limited sample}
  \label{fig:sequence}
\end{figure*}

After non-destructive examination, the sample can be analyzed with high-precision laboratory
instruments. The situation is the opposite of in-situ analysis. The instruments are highly capable,
but the sample is limited, often unique and frequently smaller than a milligram. The analytical
plan is therefore designed to obtain the most information from the least amount of material. Two
groups of techniques are most important. Mass spectrometric methods measure isotopic and elemental
composition at scales from micrometers down to near-atomic resolution. Synchrotron X-ray methods
map trace elements and chemical speciation with little damage to the sample. A single grain is
usually passed through a sequence of techniques from non-destructive to destructive, as shown in
Fig.~\ref{fig:sequence}.

\subsection{Isotopic and trace element analysis}

\Gls{sims} uses a focused primary ion beam to sputter the sample surface, and the secondary ions are
analyzed by a mass spectrometer. \Gls{sims} gives isotopic and trace element data at the micrometer
scale. \Gls{nanosims} improves the spatial resolution to about \SIrange{50}{100}{\nano\metre}
while maintaining high sensitivity. This makes it the preferred technique for the isotopic analysis
of individual sub-micrometer grains. It is especially important for presolar grains, which are
dust particles that formed around older stars before the Solar System and survived in primitive
meteorites and returned samples. Their isotopic compositions differ strongly from Solar System
material and record the nuclear processes in the stars where they formed. \citet{liu2024} reviewed
how presolar grains from supernovae are used to study stellar nucleosynthesis. \citet{nguyen2025}
studied Bennu samples and found abundant supernova dust and evidence that aqueous alteration varied
between different lithologies.

\Gls{apt} removes atoms one by one from a needle-shaped specimen using a strong electric field and
reconstructs its composition in three dimensions at near-atomic resolution. \citet{greer2020} used
\gls{apt} on space-weathered lunar ilmenite and found nanophase iron particles of about
\SIrange{3}{10}{\nano\metre}, compositional zoning and a vesicle attributed to solar wind
irradiation in the outer tens of nanometers of the grain. Such features cannot be resolved by
lower-resolution techniques. However, \gls{apt} consumes the analyzed volume, so it is placed at the
end of the analytical sequence and used on targets that have already been studied by
non-destructive methods.

\Gls{tims} remains a reference technique for dating planetary materials. In \gls{tims}, chemically
purified elements are ionized from a heated filament and isotope ratios are measured with very high
precision after correction for mass fractionation during ionization. \Gls{tims} is used for
isotopic dating systems such as uranium-lead and rubidium-strontium. Using several independent
dating systems on the same material allows the ages to be checked against each other. The same
approach is used in nuclear forensics to date nuclear materials \citep{kristo2016}.
\citet{connelly2012} used uranium-corrected lead-lead dating of calcium-aluminum-rich inclusions
and chondrules from primitive meteorites and found an age of $4567.30 \pm 0.16$ million years
for the oldest Solar System solids. They also showed that chondrules formed over about the first
3 million years of Solar System history. This level of precision can only be achieved in a
laboratory on Earth. Organic molecules in returned samples are analyzed by chromatography and mass
spectrometry. \citet{naraoka2023} identified amino acids and many other soluble organic molecules
in Ryugu samples.

\subsection{Synchrotron techniques}

Synchrotron sources produce intense, tunable and highly collimated X-ray beams that can be focused
to sub-micrometer spots. This allows two types of measurement that are especially useful for rare
samples because they cause little damage. Synchrotron micro-\gls{xrf} maps the distribution of
trace elements with high sensitivity. \Gls{xanes} spectroscopy is obtained by scanning the X-ray
energy across the absorption edge of an element, and it gives the oxidation state and local
coordination of that element. The oxidation state of iron is a common example. The energy of the
pre-edge feature in the iron K-edge spectrum changes with the average oxidation state of iron.
\citet{bajt1994} calibrated this energy against standards and used it to determine the
Fe$^{3+}$/$\Sigma$Fe ratio of individual mineral grains without destroying them. The relation can
be written as Eq.~(\ref{eq:xanes}).
\begin{equation}
  E_{\mathrm{pre}} = a\,\langle V\rangle + b
  \label{eq:xanes}
\end{equation}
Here $E_{\mathrm{pre}}$ is the centroid energy of the pre-edge feature, $\langle V\rangle$ is the
average oxidation state of iron, and $a$ and $b$ are constants determined from standards of known
oxidation state. Oxidation state and speciation can change when a sample is exposed to air and
moisture on Earth. Non-destructive mapping of pristine returned samples is therefore a major
advantage over methods that require dissolution or sectioning. The importance of avoiding
alteration is clear from the Bennu samples, in which \citet{mccoy2025} identified a sequence of
salts, including sodium carbonates, phosphates, sulfates and halides, that formed as a brine
evaporated. Many of these minerals dissolve in water and can change quickly in the terrestrial
atmosphere.

\section{Accuracy, Calibration and Reference Materials}\label{sec:accuracy}

Every measurement in the chain depends on the basic figures of merit of analytical chemistry. These
are accuracy, precision, detection limit, dynamic range and traceability to reference materials.
They determine whether a reported composition is reliable enough to support a geological or
astrobiological conclusion. In-situ instruments face the most difficult calibration problem. They
cannot analyze a fresh standard between samples. Their calibration is prepared before launch and
must remain valid for years and over large temperature changes. For \gls{libs}, the calibration
standards must also cover the full range of rock compositions expected on the planet, so that the
matrix effects described by Eq.~(\ref{eq:libs}) can be corrected \citep{clegg2017,boucher2015}.

The required accuracy depends on the scientific question. \citet{treiman2020} examined how accurate
spacecraft analyses of major elements in basalts must be to answer questions about the origin of
these rocks and compared this requirement with the performance of spacecraft instruments. Their
work shows that precision alone is not enough, because a precise but inaccurate analysis can lead
to a wrong interpretation. This is a particular concern for remote and in-situ methods because
their calibration cannot be renewed during the mission. For the same reason, contamination
knowledge (Section~\ref{sec:contam}) and matrix-matched calibration are essential parts of the
measurement.

Laboratories on Earth address these figures of merit with certified reference materials,
procedural blanks, isotope dilution, standard addition and interlaboratory comparison. Planetary
missions now apply similar practices. Examples include onboard calibration targets, witness
materials and large pre-flight calibration databases such as those used for ChemCam and SuperCam
\citep{clegg2017,maurice2021}. These practices allow in-situ and laboratory data to be compared on
a common basis. This is necessary for data fusion, because different datasets can be combined only
when their uncertainties are known and comparable.

\section{Chemometrics and Data Fusion}\label{sec:chemometrics}

The datasets described above differ in almost every way. They cover spatial scales from kilometers
to fractions of a nanometer, measure different properties such as elemental concentration,
molecular structure, isotope ratio and oxidation state, and have different calibration and
uncertainty characteristics. Combining them into one geochemical model is an analytical problem in
itself. It is addressed with the chemometric and statistical tools used throughout analytical
chemistry. For a single instrument, chemometrics converts spectra into composition. The
multivariate calibration of \gls{libs} spectra is a common example. The oxide concentrations are
related to the spectrum by a regression model of the form given in Eq.~(\ref{eq:pls}).
\begin{equation}
  \mathbf{c} = \mathbf{X}\boldsymbol{\beta} + \boldsymbol{\varepsilon}
  \label{eq:pls}
\end{equation}
Here $\mathbf{c}$ is the vector of concentrations, $\mathbf{X}$ is the matrix of spectral data,
$\boldsymbol{\beta}$ is the vector of regression coefficients obtained from standards and
$\boldsymbol{\varepsilon}$ is the residual error. \Gls{pls} regression and \gls{ica} are the main
methods used for ChemCam quantification \citep{clegg2017}. \citet{boucher2015} showed that the
choice of regression method affects the accuracy of \gls{libs} results for major elements. At the
scale of a whole sample, data fusion places datasets from different instruments and scales in a
common spatial and compositional framework. In this way, a nanometer-scale isotopic anomaly can be
related to its micrometer-scale mineralogy and millimeter-scale texture. The coordinated analysis
of Ryugu samples by \citet{nakamura2023} is an example of this approach, in which tomography,
diffraction, spectroscopy and mass spectrometry were applied to the same particles. The same idea
applies across the whole chain. The fine-scale chemical data from \gls{pixl} and SuperCam can be
compared with laboratory measurements of the same samples if they are returned in the future
\citep{allwood2020,maurice2021}.

\section{Comparison of In-Situ and Laboratory Analysis}\label{sec:comparison}

\begin{figure*}[t]
  \centering
  \includegraphics[width=0.8\textwidth]{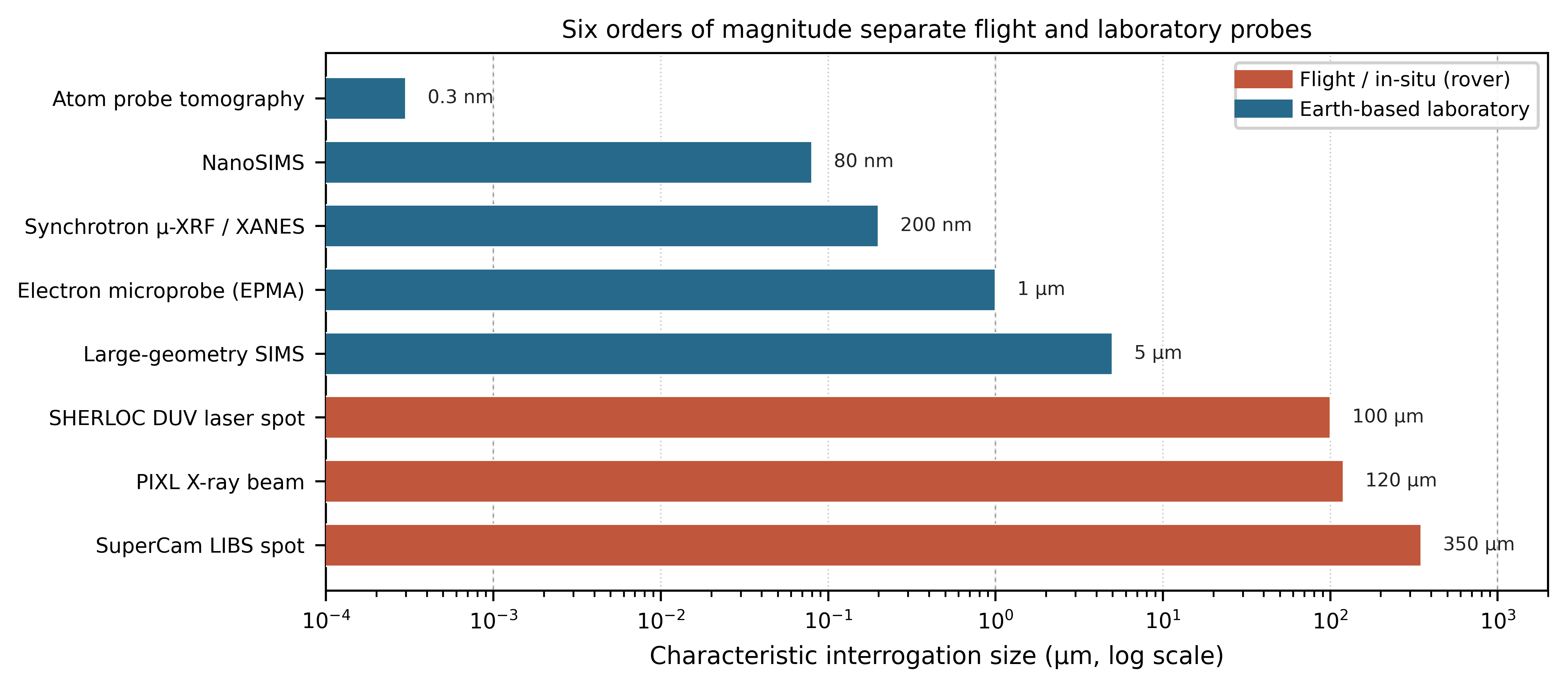}
  \caption{Characteristic analysis size of representative flight and laboratory techniques on a
  logarithmic scale. Flight instruments analyze areas about three to six orders of magnitude larger
  than the highest-resolution laboratory techniques. This difference in spatial resolution is the
  main analytical reason for returning samples to Earth}
  \label{fig:probesize}
\end{figure*}

Comparing flight and laboratory instruments on the same scale shows how they complement each other
and why sample return is valuable. The comparison is summarized in Table~\ref{tab:comparison}. The
most obvious difference is spatial resolution. The analysis spot of flight instruments is a few
hundred micrometers for SuperCam \gls{libs}, about \SI{120}{\micro\metre} for \gls{pixl} and about
\SI{100}{\micro\metre} for \gls{sherloc}. In contrast, \gls{nanosims} resolves features of tens of
nanometers and \gls{apt} approaches atomic resolution. As shown in Fig.~\ref{fig:probesize}, this
is a difference of up to six orders of magnitude. A similar gap exists in isotopic precision, which
is essential for dating and for identifying the nucleosynthetic and alteration signatures that
distinguish different parent bodies.

\begin{table*}[t]
  \centering
  \caption{Comparison of in-situ (flight) and laboratory analysis of planetary materials. Values
  indicate typical operating ranges}
  \label{tab:comparison}
  \small
  \begin{tabularx}{\textwidth}{>{\bfseries}l X X}
    \toprule
    \normalfont\textbf{Parameter} & \textbf{In-situ (flight)} & \textbf{Laboratory (Earth)} \\
    \midrule
    Typical spatial resolution  & \SIrange{100}{500}{\micro\metre} & \SI{1}{\micro\metre} down to about \SI{0.3}{\nano\metre} \\
    Elemental detection limit   & tens of ppm to wt\% (\acrshort{libs}, \acrshort{xrf}) & ppm to sub-ppb (\acrshort{sims}, \acrshort{tims}) \\
    Isotopic precision          & limited & sub-per-mil (\acrshort{tims}, \acrshort{nanosims}) \\
    Sample preparation          & none or minimal & extensive (sectioning, \acrshort{fib} milling, chemical separation) \\
    Effect on sample            & mostly non-destructive & partial or complete consumption \\
    Throughput                  & high, many targets in place & low, one grain at a time \\
    Main constraint             & mass, power, data volume, matrix effects & limited and often unique sample mass \\
    Representative instruments  & SuperCam, \acrshort{pixl}, \acrshort{sherloc}, \acrshort{sam} & \acrshort{nanosims}, \acrshort{apt}, \acrshort{tims}, synchrotron micro-\acrshort{xrf} \\
    Example result              & organic matter detected in situ (\acrshort{sam}) & amino acids, nucleobases and ammonia (Ryugu, Bennu) \\
    \bottomrule
  \end{tabularx}
\end{table*}

\begin{figure*}[t]
  \centering
  \includegraphics[width=0.8\textwidth]{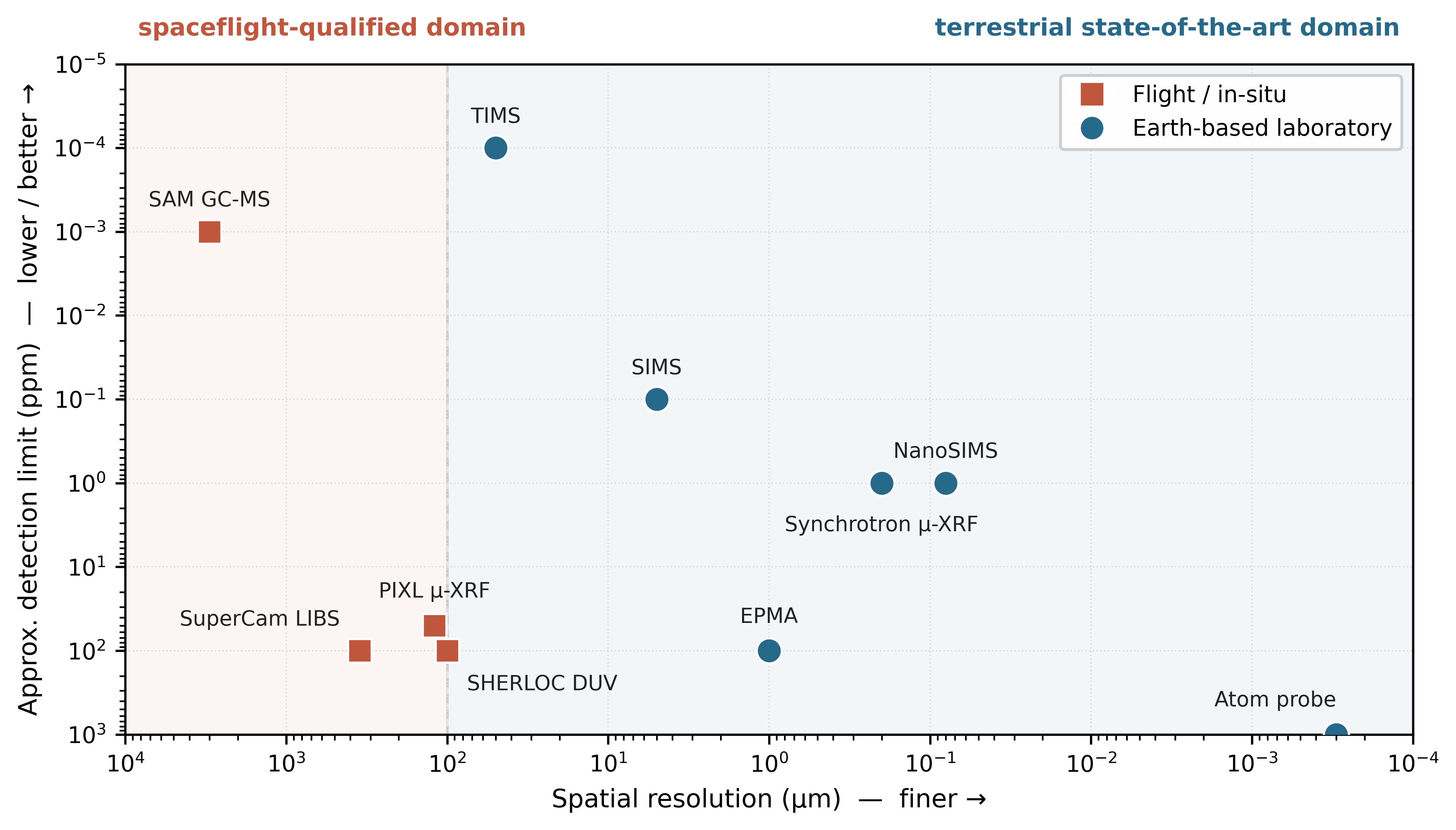}
  \caption{Approximate positions of planetary analytical techniques in terms of spatial resolution
  and detection limit. Flight instruments (squares) and laboratory instruments (circles) occupy
  mostly separate regions}
  \label{fig:capability}
\end{figure*}

Spatial resolution and detection limit can be combined in one plot, as shown in
Fig.~\ref{fig:capability}. Flight and laboratory instruments occupy mostly separate regions of this
plot. Flight instruments are designed for wide coverage, autonomous operation and non-destructive
analysis of unprepared surfaces. In return, they accept lower spatial resolution, and, in the case
of \gls{libs}, quantification limited by matrix effects. Laboratory instruments are designed for
high spatial resolution, precision and speciation, but they require a physical sample and often
consume part or all of it. The study of organic matter illustrates this difference. \Gls{sam}
detected organic matter at Gale crater but could determine its molecular structure only partly
because of pyrolysis \citep{eigenbrode2018}. In contrast, laboratory analysis of returned asteroid
samples has identified amino acids, nucleobases, ammonia and nitrogen-rich soluble organic
compounds in contamination-controlled material \citep{naraoka2023,glavin2025}.

This comparison should not be read as a ranking. In-situ instruments provide reconnaissance and
context, which are needed to select the right samples. Decisions such as which rock to sample and
which part of the rock to analyze can only be made at the site. Laboratory instruments then obtain
isotopic, molecular and speciation data from the selected samples that no flight instrument can
provide at present. The precise chronology of the early Solar System \citep{connelly2012} and the
nanometer-scale records of stardust and space weathering in returned grains
\citep{nguyen2025,greer2020} can at present be obtained only in laboratories on Earth. These
results also depend on the in-situ selection of samples and on contamination control during
collection and curation.

\begin{figure*}[t]
  \centering
  \includegraphics[width=0.72\textwidth]{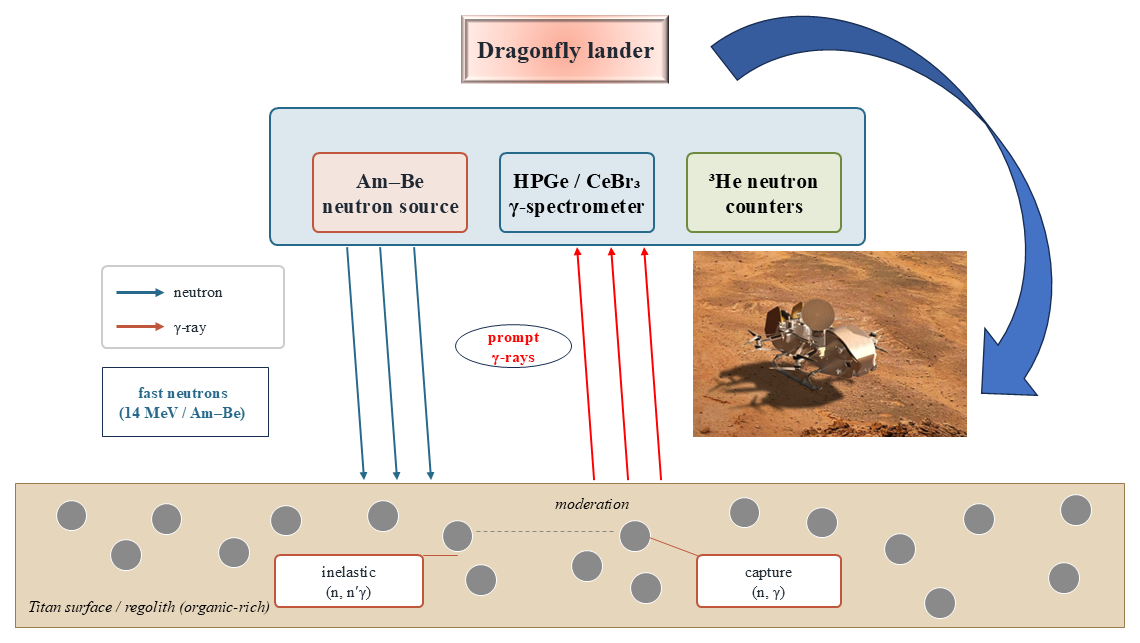}
  \caption{Schematic of active neutron-gamma analysis by \acrshort{dragns} on Titan. Titan's dense
  atmosphere blocks the galactic cosmic rays used by passive orbital gamma-ray and neutron
  spectroscopy, so \acrshort{dragns} provides its own fast neutrons from a pulsed neutron
  generator (original design) or an americium-beryllium source (current design). The neutrons
  undergo inelastic scattering and thermal neutron capture in the surface material. The resulting
  prompt gamma rays, which are characteristic of each element, are measured by a passively cooled
  \acrshort{hpge} detector with a CeBr$_3$ scintillator as backup, while $^{3}$He counters measure
  the thermal and epithermal neutrons. Dragonfly rotorcraft illustration: NASA/Johns Hopkins APL}
  \label{fig:dragns}
\end{figure*}

\section{Active Neutron-Gamma Spectroscopy on Titan}\label{sec:titan}

The in-situ techniques discussed in Section~\ref{sec:insitu} are best suited to Mars. On other
bodies, different methods are needed. An important upcoming example is the \gls{dragns} on NASA's
Dragonfly mission to Saturn's moon Titan. Dragonfly is a rotorcraft lander that will fly between
many sites in the organic-rich Shangri-La dune region to study prebiotic chemistry
\citep{barnes2021}. NASA confirmed the mission in April 2024 with a planned launch in July 2028 and
arrival at Titan in 2034 \citep{nasa2024}. Titan is difficult for the methods used on Mars. Its
surface temperature is about \SI{94}{\kelvin}, and its nitrogen atmosphere is about four times
denser than Earth's atmosphere at the surface \citep{barnes2021}. This thick atmosphere prevents
galactic cosmic rays from reaching the surface. As a result, the passive gamma-ray and neutron
spectroscopy described in Section~\ref{sec:orbital} cannot be used, and \gls{dragns} must provide
its own neutrons \citep{lawrence2022}. The measurement principle is shown in Fig.~\ref{fig:dragns}.

This technique is the planetary form of \gls{pgnaa}, which is used on Earth for bulk material
analysis and in borehole logging. The original \gls{dragns} design used a deuterium-tritium pulsed
neutron generator that produces \SI{14}{\mega\electronvolt} neutrons at about $10^{8}$ neutrons per
second \citep{lawrence2022}. The current design uses an americium-beryllium radioactive source
called \gls{fire}, which reduces systematic uncertainties and resource requirements while keeping
the science objectives \citep{wilson2025}. The neutrons enter the near-surface material, where they
undergo inelastic scattering and thermal neutron capture. Each reaction emits prompt gamma rays
with energies that identify the element and intensities that depend on its concentration. Hydrogen
and nitrogen are identified by capture lines at \num{2.223} and \SI{10.829}{\mega\electronvolt},
while carbon and oxygen are identified by inelastic scattering lines at \num{4.439} and
\SI{6.130}{\mega\electronvolt}. The gamma rays are measured with a \gls{hpge} detector that is
cooled passively by Titan's cold atmosphere rather than by a mechanical cooler. Its energy
resolution of about 0.3\% at \SI{1333}{\kilo\electronvolt} is much better than the approximately
3\% of the best scintillators. A CeBr$_3$ scintillator provides backup, and two $^{3}$He counters
measure thermal and epithermal neutrons \citep{lawrence2022}. The use of an americium-beryllium
source and the moderation of fast neutrons in hydrogen-rich material follow the same principles as
laboratory \gls{naa}.

Within the analytical chain, \gls{dragns} plays the same role on Titan that \gls{libs} and
\gls{xrf} play on Mars. It measures bulk carbon, hydrogen, nitrogen and oxygen, which are
indicators of prebiotic chemistry, as well as phosphorus, sulfur and the salt-forming elements
sodium, chlorine and potassium \citep{lawrence2022}. It does this without drilling, contact or
sample preparation. The measurements can be made at each landing site and help the team decide
where the drill and the Dragonfly Mass Spectrometer should carry out more detailed molecular
analysis \citep{barnes2021}. \Gls{dragns} gives up spatial and molecular detail in exchange for
depth-integrated elemental data that do not consume the sample. It therefore complements the mass
spectrometer in the same way that in-situ survey complements laboratory micro-analysis elsewhere in
the chain. The use of \gls{naa}, which has long been a laboratory technique, on a mission to Titan
shows how far planetary analytical chemistry has extended beyond optical and mass spectrometric
methods.

\section{Conclusions}\label{sec:conclusions}

This review has shown that the analytical chemistry of planetary science can be understood as one
connected chain. The most valuable measurements require the most capable instruments, which are
located on Earth. However, the materials of interest are located on other bodies and reach the
laboratory, if at all, in small and irreplaceable amounts. Each part of the chain responds to this
problem in a different way. Orbital methods produce compositional maps that guide where to land
and sample. In-situ instruments give up spatial resolution and precision in return for the ability
to analyze samples in place, without destroying them and with little human intervention.
\Gls{libs}, \gls{xrf}, \gls{duv} Raman, evolved gas analysis and miniaturized mass spectrometry now
routinely provide elemental, mineralogical and molecular data from the Martian surface
\citep{wiens2021,allwood2020,bhartia2021,mahaffy2012}. Contamination control and curation have
become quantitative disciplines, and blanks, witness materials and cleanliness requirements are
necessary for interpreting any returned sample \citep{farley2020,maltais2023}. Laboratory
techniques provide isotopic, chronological and speciation data from sub-milligram grains at scales
down to near-atomic resolution. Calibration, reference materials and uncertainty estimation connect
the in-situ and laboratory data, and chemometric data fusion combines them into geochemical models.
The coming years will place new demands on this chain. The Ryugu and Bennu samples have already
moved the field from meteorites of uncertain origin to pristine samples of known origin and
context. These samples have revealed prebiotic organic compounds and records of water-rock
interaction and brine evaporation. The samples collected by Perseverance remain on Mars without an
approved return mission after the termination of the NASA \gls{msr} program. If Martian samples are
returned in the future, they will require the strictest contamination control, non-destructive
preliminary examination and coordinated micro-analysis, because their astrobiological value
depends on separating Martian chemistry from terrestrial contamination at trace levels. The main
tasks for the next decade are therefore to improve the accuracy of in-situ instruments, reduce
contamination levels and develop coordinated analytical and data fusion workflows. These steps will
allow a single returned grain to be studied from its three-dimensional structure down to individual
atoms with minimum loss of material.

\printnoidxglossary[type=\acronymtype,title={Abbreviations},sort=word]

\section*{Acknowledgements}
The author thanks the Department of Physics, University of Idaho, for providing the academic
environment and support in which this work was carried out. The author also thanks the University
of Idaho Library for access to the journals, databases and other academic resources used in
preparing this review.

\section*{Funding}
This work did not receive any specific grant from funding agencies in the public, commercial or
not-for-profit sectors.

\section*{Author contributions}
M.A. conceived the review, carried out the literature search and analysis, prepared the figures
and tables, and wrote and revised the manuscript. The author read and approved the final
manuscript.

\section*{Data availability}
This review does not report new experimental data. All data discussed are available in the
published sources cited in the reference list. The values used to construct the figures were
compiled from these sources and are available from the author on reasonable request.

\section*{Competing interests}
The author declares no competing interests.

\section*{Third-party material}
All figures and tables were created by the author for this review, except for the Dragonfly
rotorcraft illustration in Fig.~\ref{fig:dragns}, which is credited to NASA/Johns Hopkins APL.

\bibliographystyle{plainnat}
\bibliography{references}

\end{document}